# WawPart: Workload-Aware Partitioning of Knowledge Graphs

Amitabh Priyadarshi and Krzysztof J. Kochut

University of Georgia, Athens GA 30602, USA
{amitabh.priyadarshi, kkochut}@uga.edu

**Abstract.** Large-scale datasets in the form of knowledge graphs are often used in numerous domains, today. A knowledge graph's size often exceeds the capacity of a single computer system, especially if the graph must be stored in main memory. To overcome this, knowledge graphs can be partitioned into multiple sub-graphs and distributed as shards among many computing nodes. However, performance of many common tasks performed on graphs, such as querying, suffers, as a result. This is due to distributed joins mandated by graph edges crossing (cutting) the partitions. In this paper, we propose a method of knowledge graph partitioning that takes into account a set of queries (workload). The resulting partitioning aims to reduces the number of distributed joins and improve the workload performance. Critical features identified in the query workload and the knowledge graph are used to cluster the queries and then partition the graph. Queries are rewritten to account for the graph partitioning. Our evaluation results demonstrate the performance improvement in workload processing time.

## 1 Introduction

In today's world, the data or information is interconnected and form large knowledge graphs or information networks. Such knowledge graphs are often used to represent data in social networking systems, shopping and movie preferences, bioinformatics, and in other real-world systems. The availability of large-scale data, represented as knowledge graphs composed of millions or even billions of vertices and edges, requires large-scale graph processing systems. Such data is often too large to be stored at one place in a centralized data store and needs to be partitioned into multiple sub-graphs, often called shards, and transferred to multiple computing nodes in a distributed system. However, with the overall knowledge graph split into distributed shards, many typical graph-base tasks, such as query processing, suffer from network latency and other problems related to the original graph being partitioned. One of the techniques to improve query processing performance in such a system is to reduce the inter-node communication between graph processing systems. The graph partitioning can be an effective pre-processing technique to balance the runtime performance.

In this paper, we will assume that a knowledge graph is a dataset represented using the Resource Description Framework (RDF) [1]. An RDF knowledge graph is a set of RDF triples in a form of (*s*, *p*, *o*), where *s* is the subject, *p* predicate, and *o* the object. A predicate represents a semantic relationship connecting a subject and an object. Consequently, an RDF dataset is a directed graph, where nodes (vertices) and edges have types, often described using the Resource Description Framework Schema (RDFS) [2]. In many recent publications, a knowledge graph has often been defined as a Heterogeneous Information Network (HIN) [3], a form of a directed graph, where nodes and edges have heterogeneous types. HINs are a bit more restrictive than RDF/RDFS in that if two edges are labeled by the same relationship type, their starting and ending nodes must have the same object types, respectively. RDF/RDFS does not have this restriction. Recently, graph databases, such as Neo4j [4], have also been used to represent and store knowledge graphs.

Knowledge graphs have been used for many different tasks, such as data mining, link (edge) prediction, and data classification, to mention a few. Query processing, intended to retrieve some data of interest, is one of the most common tasks. The SPARQL query language is the query language for RDF/RDFS datasets [5], while Cypher is an example graph query language used with the popular Neo4j graph database [4]. A critical part of a SPARQL query is a graph pattern composed of *triple patterns* involving variables. Without going into details here, a query processor matches the query pattern against the knowledge graph and reports matches as query solutions.

Typically, graph tasks become computationally intensive, when considering graphs with millions, even billions of vertices. A large knowledge graph can easily overload the memory and processing capacity even on relatively large servers, especially for in-memory knowledge graph systems. Partitioning a knowledge graph into smaller components (sub-graphs) and distributing the partitions, called shards, across a cluster of servers may enable handling of very large graphs and potentially offer performance improvements in executing different tasks.

Given a graph $G = (V, E)$, where *V* is a set of vertices and *E* is a set of edges and a number $k > 1$, a *graph partitioning of G* is a subdivision of vertices of *G* into *subsets* of vertices $V_1, \ldots, V_k$ that partition the set V. A *balance constraint* requires that all partition blocks are equal, or close, in size. In addition, a common objective function is to minimize the *total number of cuts*, i.e., edges crossing (cutting) partition boundaries.

This paper is outlined as follows. Section 2 provides an overview of related work. Section 3 discuss about the partitioning method. Section 4 is about the Architecture the experiments, and section 5 concludes the paper.

## 2  Related Work

It is generally expected that partitioning a large-scale graph decreases query processing efficiency. However, this decrease can be moderated if the partitioning is sensitive to a query workload used on a daily basis and optimized to minimize the inter-partition communication needed by the workload. In this section, we discuss research results related to graph partitioning and their impact on querying.

The graph partitioning is an NP-complete [9] problem. To alleviate this problem, many practical solutions have been developed, including spectral partitioning methods [10] and geometric partitioning methods [11]. The multilevel method was introduced to the graph partitioning by Barnard and Simon [6] and then improved by Hendrickson and Leland [7]. The multilevel method consists of three main phases: coarsening, initial partitioning, and uncoarsening. A partitioning approach of Karypis et al [8] uses recursive multilevel bisection method for bisection of a graph to obtain a k-partition on the coarsest level.

ParMETIS [12] and LogGP [13] are state-of-the-art parallel graph processing systems. Also, LogGP is a graph partitioning system that records, analyzes, and reuses the historical statistical information to generate a hyper-graph and uses a novel hyper-graph streaming partitioning approach to generate a better initial streaming graph partition. Pregel [14] is used for graphs with billions of vertices and is built upon MapReduce [15], and its open-source Apache Giraph [16], PEGASUS [17] and GraphLab [18] used as a default partitioner of vertices. Sedge [19] implements a similar technique as Pregel, which is also a vertex-centric processing model for SPARQL query execution implemented on top of this model. These methods, however, do not take into account a set of queries, called a workload, executed against the partitioned graph. This requires distributed query processing which penalizes workload processing performance.

DREAM [20], WARP [21], PARTOUT [22], AdPart [23], and WISE [24] are workload-aware, distributed RDF systems. DREAM [20] partitions only SPARQL queries into subgraph patterns and not the RDF dataset. The RDF dataset is replicated among nodes. It follows a master-slave architecture, where each node uses RDF-3X [25] on its assigned data for statistical estimation and query evaluation. WARP [21] uses the underlying METIS system to assign each vertex of the RDF graph to a partition. Triples are then assigned to partitions, which are stored at dedicated hosts in a triple store (RDF-3X). WARP determines a query's center node and radius based on an n-hop distance. If the query is within n-hops, WARP sends the query to all partitions to be executed in parallel. A complex query is decomposed into several sub-queries and executed in parallel and then the results are combined. PARTOUT [22] extracts representative triple patterns from a query workload by applying normalization and anonymization by replacing infrequent URIs and literals with variables. Frequent URIs (above a frequency threshold) are normalized. PARTOUT uses an adapted version of RDF-3X as a triple store for their n hosts. AdPart [23] is an in-memory RDF system that re-partitions the RDF data incrementally. It uses hash partitioning that avoids the cost associated with initial partitioning of the dataset. Each worker stores its local set of triples in an in-memory data structure. AdPart provides an ability to monitor and index workloads in the form of hierarchical heat maps. It introduces Incremental ReDistribution (IRD) technique for data portions that are accessed by hot patterns. IRD is a combination of hash-partitioning and k-hop replication, guided by a query workload. WISE [24] is a runtime-adaptive workload-aware partitioning framework for large scale knowledge graphs. A partitioning can be incrementally adjusted by exchanging triples, based on changes in the workload. SPARQL queries are stored in a Query Span structure with their frequencies. To reduce communication overhead, the system redistributes frequent query patterns among workers. A cost model which maximizes the migration the gain while preserving the balanced partition is used for migration of triples.

Our query workload-aware knowledge graph partitioning method, called WawPart, which extracts critical features from the query workload as well as from the dataset. These features are used to establish distance among queries in a form of distance matrix and then cluster similar queries together using hierarchical agglomerative clustering. Subgraphs (partitions) associated with these features are then created from the knowledge graph data and distributed as shards in a computing cluster.

The five closely related systems discussed in the previous paragraph rely on specialized RDF/SPARQL processing systems. In contrast, these systems, ours does not rely on a specialized data store implementation and uses an *off-the-shelf* knowledge graph storage and query processing system (Virtuoso [28]) and relies on standard SPARQL queries for distributed processing. We believe that it is an important advantage.

## 3   Workload-Aware Knowledge Graph Partitioning

Knowledge graph partitioning created by WawPart attempts to optimize query workload processing time. Critical features of both the workload queries and the knowledge graph are extracted and analyzed. Subsequently, the queries are clustered based on their similarity and the knowledge graph is then partitioned based on the clustering. Original queries are re-written into federated SPARQL queries, since partitions are distributed among computing nodes. Typically, query processing times are increased, due to the distributed joins, as connected triples may be stored in partitions distributed to different nodes (such distributed partitions are usually called *shards*.) In this paper we focus on the effects of workload-aware knowledge graph partitioning, given a query workload. Here, we do not consider the changes in the workload and necessary modifications to the partitioning.

### 3.1   Query and Knowledge Graph Feature Extraction

Searching for similarities among query graph patterns is computationally expensive, as finding overlaps among query graph patterns is similar to finding isomorphic subgraphs. To help with these problems, we identify and extract critical features of graph patterns in a workload. Similarly, we identify critical features in the knowledge graph (later, we refer to it as the dataset, as well). These features are used to analyze the similarity among triples in queries. We identify the following features in the workload query graph patterns and in the knowledge graph:
• *Predicate (P)* feature represents all triples that share a given predicate (edge label).
• *Predicate-Object (PO)* represents all triples that share a given predicate *and* an object value. This feature is useful in analyzing similarity of some queries.

Other features may be useful in query analysis, as well. These include:
- *Subject-Subject (SS)* feature represents triples that share the same subject, i.e., edges sharing the same source vertex. This feature is used for analyzing queries involving multiple predicates/edges with the same subject (often called star shape patterns).
- *Object-Subject (OS)* represents triples where one triple's object (destination vertex) is another triple's subject (source vertex). Such data representation is beneficial when a query includes pairs of triples with connection based on object - subject ( sometimes called an elbow join).
- *Object-Object (OO)* feature represents triples that share the same object (target).

Triples identified by the SS, OS, and OO features represent joins on connecting vertices, which are shared entities (RDF resources) in the knowledge graph. Typically, query patterns also involve joins based on variables shared between triple patterns. All features extracted from the workload queries and from the knowledge graph are stored as metadata. We use them to drive the knowledge graph partitioning. To extract the features from these queries, we use a query analyzer, which analyzes the entire workload. Our query analyzer has been tailored for SPARQL queries, but it can be easily modified for other graph pattern-based query languages and knowledge graph representations, such as the Cypher language used in the Neo4j graph database [4].

In order to speed up handling of the features in the knowledge graph, we created indices on all triples in the knowledge graph, based on their subjects, predicates and objects. We used Apache Lucene API [26] to create the indices and to quickly materialize the necessary features in the knowledge graph. For example, it is possible to materialize the *Predicate (P)* feature and find all triples using a given predicate or materialize the *Predicate-Object (PO)* and find triples sharing a given predicate *and* object. Selected triples can then be easily assigned to their intended partitions, as needed.

### 3.2 Query Workload Clustering and Knowledge Graph Partitioning

We measure similarity between queries based on their features, as explained above. A distance matrix is often used as the basis for many data mining tasks, such as multi-dimensional scaling, hierarchical clustering, and others. We use Jaccard similarity to construct the distance matrix for the graph patterns in the workload queries. The Jaccard

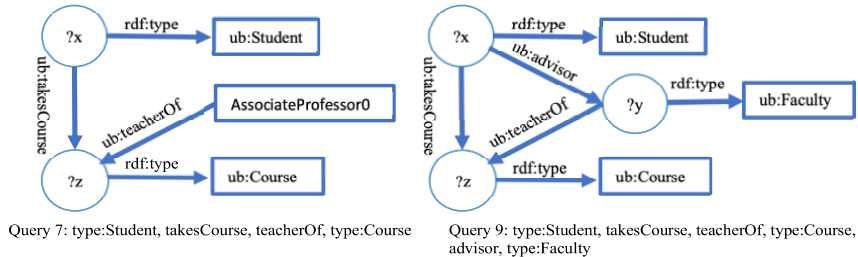

Query 7: type:Student, takesCourse, teacherOf, type:Course    Query 9: type:Student, takesCourse, teacherOf, type:Course, advisor, type:Faculty similarity of sets A and B is the ratio of the intersection of set A and B to the union of set A and B, or $J_{sim}(A,B) = |A \cap B| / |A \cup B|$, while the Jaccard distance between sets A and B is given by $1 - J_{sim}(A,B)$. Intuitively, if graph patterns in two queries are nearly identical, based on their features, the distance between them is zero or close to zero. We use a similarity matrix to represent the similarity of queries in the workload and use the matrix to split the workload queries into subsets of similar queries. The details of the computation are skipped here. As an example, consider 2 queries in Fig. 1. Query 7 has 4 features: (2 *PO* features: *rdf:type*→ *ub:Student, rdf:type*→ *ub:Course* and 2 *P* features: *ub:takesCourse, ub:teacherOf*) while query 9 has 6 features (3 *PO* features: *rdf:type*→ *ub:Student, rdf:type*→ *ub:Course, rdf:type*→ *ub:Faculty* and 3 *P* features: *ub:takesCourse, ub:teacherOf, ub:advisor*). Consequently, the distance between these queries is $(1 - J_{sim}) = 1 - |Q7 \cap Q9| / |Q7 \cup Q9| = 0.33$.

**Fig. 1.** Distance between Q.7 and Q.9 is 1- $J_{sim}$ = 1- $|Q7 \cap Q9|$ / $|Q7 \cup Q9|$ = (1- 4/6) = 0.33.

| | **Algorithm 1.** Hierarchical Agglomerative Clustering of Query Workload |
|---|---|
| Input | Feature Distance Matrix ***D*** of workload Query. |
| Output | HAC Dendrogram I |
| 1 | Assign for each D[n][n] into C[m] where $C = c_1, c_2, …, c_m$ |
| 2 | **while** C.size > 1 **do** |
| 3 |    **for** i = 1 **to** C.size **do** |
| 4 |       **if** $(c_a, c_b) = min\ d(c_a, c_b)$ in C **then** *// distance function $d(c_1, c_2)$* |
| 5 |          delete $c_a$ and $c_b$ from C |
| 6 |          add $min\ d(c_a, c_b)$ in C |
| 7 |       assign **I** = (old, $c_a c_b$,, $min\ d(c_a, c_b)$)) *//(oldGroup, newGroup, distance)* |
| 8 |       recalculate proximity matrix using (SL/CL/AL)<br>         **P = modifyD(** $c_a, c_b, min\ d(c_a, c_b)$**))** *//(oldGroup, newGroup, distance)* |
| 9 |    for each P, $c_m$ = P[i][j] |
| 10 |    Update $C = c_1, c_2, …, c_m$ |
| 11 | Output **I** |

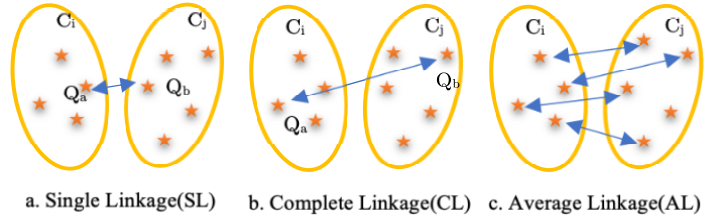

a. Single Linkage(SL)    b. Complete Linkage(CL)    c. Average Linkage(AL)

**Fig. 2.** Linkages **a)** $SL(C_i, C_j) = min(D(Q_a, Q_b))$, **b)** $CL(C_i, C_j) = max(D(Q_a, Q_b))$ and **c)**
$$AL(C_i, C_j) = \frac{1}{n_{Ci} n_{Cj}} \sum_{a=1}^{n_{Ci}} \sum_{b=1}^{n_{Cj}} D(Q_a, Q_b)$$

For cluster analysis of the features in queries, we used the hierarchical agglomerative clustering (HAC), which is a method of creating a hierarchy of clusters in a bottom-up fashion (Algorithm 1). The decision as to which features should be kept together is established by HAC based on the measure of similarity between queries and selection of a linkage method. The cluster grouping is done according to the shortest distance among all pairwise distances between queries. Once the two most similar queries are grouped together, the distance matrix is recalculated. The first similarity measure is provided by the initial distance matrix. Recalculation of the distance matrix is based on the choice of the linkage from single, complete, or average. Fig. 2. shows linkages in which a single linkage is the proximity between two nearest neighbors where a complete linkage is the proximity between the farthest neighbor. These steps are repeated until there is only a single cluster left. We ran HAC using a single linkage against queries to obtain a dendrogram which is then used to partition the dataset. This method of partitioning reduces the inter-partition communication and reduces

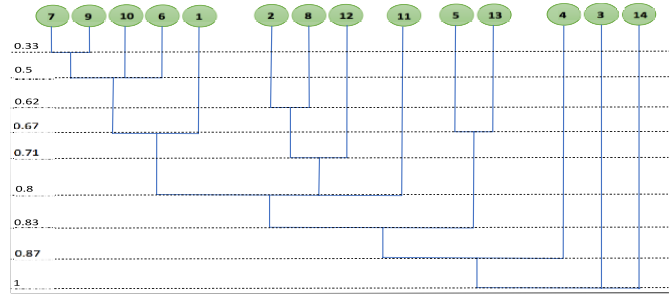

**Fig. 3.** HAC Dendrogram of LUBM's 14 Queries

distributed joins in federated queries.

The partitioning algorithm (Algorithm 2) takes a dendrogram created for a given workload and outputs a partitioning optimized for the workload. For example, we used HAC with a single linkage on the LUBM queries to produce a dendrogram shown in Fig. 3. A statistics module of the algorithm calculates the score for each replicated feature in every partition. These scores are based on 1) features which move together or are tightly connected with a replicated feature, 2) the size of the replicated feature and its peers, 3) dependencies of other queries on these features (replicated and their peer features) and 4) the number of distributed joins, when involved replicated features is on another partition. This score helps the algorithm to determine where to place the replicated features. The balancing module of the algorithm uses these features and also features that are not involved in any workload but present in the dataset to balance the partitions.

| | **Algorithm 2.** Knowledge Graph Partitioning Algorithm |
|---:|:---|
| Input | HAC Dendrogram **I**, Features **F$_G$** |
| Output | Partition metadata **P** |
| 1 | Create Feature set **g** based on **I** at similarity distance **d** |
| 2 | **Statistics(g, F$_Q$)**  //All Features *F$_G$* = (Queries features *F$_Q$*+ Unused features *F$_X$*) |
| 3 | Find **F$_R$** replicated features in **g**. |
| 4 | Find Query distributed joins of replicated features **D$_{QR}$** |
| 5 | Find stats **S$_R$** ∀ **F$_R$** |
| 6 | Find p, q and c for C and T. |
| 7 | **S$_R$** = Σ ( p$_c$w$_1$+ q$_c$w$_2$ + s$_c$w$_3$) + (p$_t$w$_4$ + q$_t$w$_5$ + s$_t$w$_6$) |
| | //**p**(Peer Features), **q**(Features in query) and **s**(Feature data size), w(weights), C(shards) and T(dataset). |
| 8 | **score for each F$_R$ = (D$_{QR}$ ∗ $w_7$)+ S$_R$** |
| 9 | **Balance_Partition(score, g , F$_G$)** |
| 10 | Remove all **F$_R$** from sets of **g** with lower **scores** for each **F$_R$** |
| 11 | Assign **data** associated to features set **g** into **P** |
| 12 | **Proximity_Query()** |
| 13 | Find proximity $F_{prox}$ = proximity of $F_{Unclustered}$ with $F_{Clustered}$. // $F_Q = F_U + F_C$. |
| 14 | Assign max($F_{prox}$) in cluster $P_i$ where its neighbor features are. |
| 15 | Assign $F_X = F_X$ + remaining $F_U$ |
| 16 | **while** $F_X$ not empty **do** |
| 17 | $P_{min}$ = Find min($P_i$) by size of shard |
| 18 | $F_{max}$ = Find max(F in $F_X$) by size/count of triples |
| 19 | Assign $F_{max}$ into $P_{min}$ |
| 20 | Output **P** |

Queries are rewritten as federated SPARQL queries [27] to accommodate the distributed shards. The overall objective is to limit the number of distributed joins in federated queries. The query rewriter computes a cost-effective query execution plan by analyzing the metadata, converts the query into a federated SPARQL query and sends it to the most suitable processing node, where the query is executed. The SERVICE keyword followed by a SPARQL endpoint directs a part of the query for processing by SPARQL endpoint on remote shard. Table 1 shows an example of an original and a rewritten, (federated) query. If all data needed for a query is present in the same shard, the query is not rewritten. If the needed triples are in different shards, we need a federated query to obtain the required data from different shards, in order to produce the final result.

Table 1. Original and Federated query of LUBM 2nd Query.

| Original Query | Federated Query |
|---|---|
| SELECT ?X ?Y ?Z FROM <lubm><br>WHERE<br>{<br>  ?X rdf:type ub:GraduateStudent .<br>  ?Y rdf:type ub:University .<br>  ?Z rdf:type ub:Department .<br>  ?X ub:memberOf ?Z .<br>  ?Z ub:subOrganizationOf ?Y .<br>  ?X ub:undergraduateDegreeFrom ?Y<br>} | SELECT ?X ?Y ?Z FROM <lubm><br>WHERE<br>{<br>  ?X rdf:type ub:GraduateStudent .<br>  ?Y rdf:type ub:University .<br>  ?Z rdf:type ub:Department .<br>  SERVICE <http://172.19.48.185:8890/sparql> {?X ub:memberOf ?Z .}<br>  SERVICE <http://172.19.48.185:8890/sparql> {?Z ub:subOrganizationOf ?Y .}<br>  SERVICE <http://172.19.48.183:8890/sparql> {?X ub:undergraduateDegreeFrom ?Y}<br>} |

## 4 Experiments and Evaluation

We have implemented a prototype system which stores an RDF knowledge graph by partitioning it into shards and distributing them among computing nodes. Our system is deployed on cluster with a Master node and separate triple stores (Virtuoso [28]) are on shared-nothing nodes. The master node is responsible for querying and getting results along with tracking execution time. Each node with a triple store is called a Processing Node, as shown in the Fig. 4. In the context of distributed RDF stores, the triples of the knowledge graph are partitioned into shards and assigned to different processing nodes with the help of the Partition Manager. The Query Rewriter/Processor (QRP) rewrites the incoming query into a federated query, according to the current location of features. The new federated query is executed on the specific shard (partition) with a maximum number of features to minimize the need for distributed joins. The shard where the query is executed is called Primary Processing Node (PPN) for that specific federated query. The system returns the results of that query to the user.

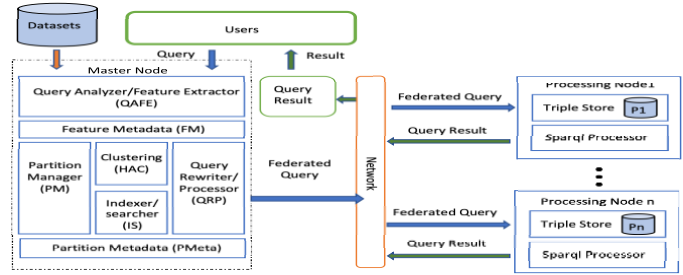

**Fig. 4.** System Architecture

In order to evaluate our knowledge graph partitioning method based on a query workload we used two synthetic datasets and associated data generation: Lehigh University Benchmark (LUBM) [29] and Berlin SPARQL Benchmark (BSBM) [30]. Both datasets were created for the purpose of testing and benchmarking Semantic Web triple store systems and their SPARQL query processing. LUBM includes a data generator, which creates synthetic OWL datasets with basic data about universities, and a set of 14 SPARQL queries for evaluation purposes. BSBM represents data about e-commerce, in which vendors offer products to customers and customers review those products. BSBM provides a dataset generator and a set of 12 SPARQL queries.

We used a cluster of Intel i5 based systems running Linux Ubuntu 18.04.4 LTS 64-bit operating system. We used relatively small machines to observe the effects of partitioning on manageable sizes of datasets. Each computer node had an instance of Virtuoso Open Source 7.2, which we used as a triple store and query processor for the triple shards. The knowledge graph partitioning systems and the experiments were coded in the Java programming language with the use of Apache Jena framework [31].

### 4.1 Results and Evaluation

Experiment shows the improvements of our workload-aware partitioning method over random partitions (complete sets of all triples with the same predicate were randomly assigned to partitions). For this experiment we have chosen LUBM dataset of 10 universities with 1,563,927 triples and BSBM dataset of 1000 products with 374,911 triples. The experiment shows the workload-aware partitioning has a significant improvement in the query processing time over a random partition.

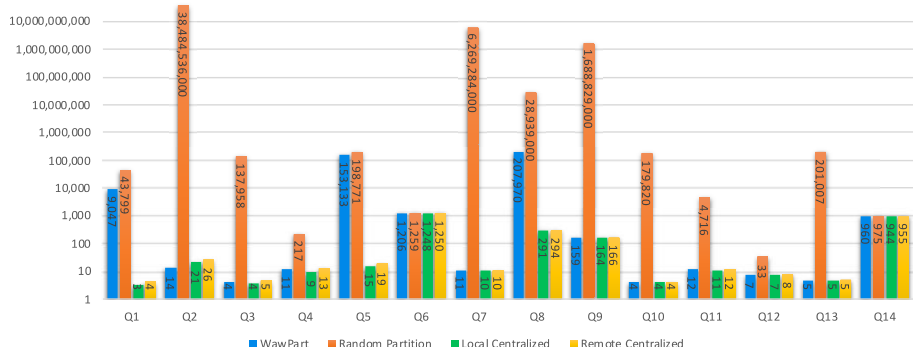

**Fig. 5.** LUBM 14 queries runtime in milliseconds.

Our partitioning algorithm created shards of the LUBM dataset, which were then distributed among the nodes in our test cluster. As an example of balancing the shard sizes, WawPart splits the LUBM's 1,564k triples into three shards of 481k, 481k and 600k triples, which is within -8% to +15% of the exact average shard size. Random Partition has shards with 521k triples each. This experiment uses two distributed and two centralized datasets and demonstrates the improvement of query workload runtime performance on a distributed system. It also shows the comparison against the baseline query workload runtime. For distributed triple datasets, we used WawPart's and randomly partitioned datasets, marked as *Random Partition*. To compute network latency, we used difference in runtime between centralized datasets, one marked as *Local Centralized* and another marked as *Remote Centralized*. Fig. 5 shows the runtime of all 14 LUBM queries in milliseconds. For 12 twelve queries (out of 14), performance of WawPart exceeded the *Random Partition* For two queries, 6 and 14, performance was the same because both queries include only a single triple pattern. In fact, performance of all queries in WawPart is very close to *Local Centralized*, which is a centralized dataset. Fig. 7 shows the average runtime of all queries and it demonstrates a dramatic improvement of WawPart's (26 second versus roughly 38 days *on Random Partition*). The average runtime for centralized datasets 0.2 seconds. For the BSBM dataset, Fig. 6 shows the query runtime for all 12 queries. Fig. 8 shows the average runtime of all 12 queries, where we can see the improvement of WawPart against the *Random Partition*. The distributed random partition average runtime is 1 hour 49 minutes and WawPart average runtime is only 16.7 milliseconds. The base average runtime is around 12.9 milliseconds for *Local All* in the centralized dataset. Performance of WawPart (distributed) is almost always very close to the centralized *Local All* partitioning.

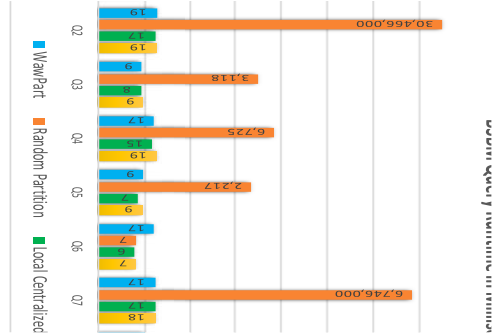

**Fig. 6.** BSBM 12 queries runtime in milliseconds.

Our experiments demonstrate that partitioning the dataset without workload awareness significantly decreases the query workload performance since distributed joins are expensive. Partitioning the knowledge graph takin into account a given query workload leads to significant performance gains. In contrast these systems, ours does not rely on a specialized data

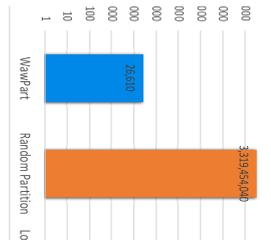
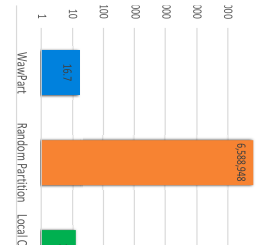

**Fig. 7.** LUBM 14 queries average runtime  **Fig. 8.** BSBM 12 queries average runtime

## 5  Conclusions and Future Work

In this paper, we have proposed a WawPart system, which is a knowledge graph partitioning and query processing system. It partitions the RDF dataset based on the query workload and aims to reduce the number of distributed joins during query execution, to improve the workload run-time. WawPart requires no replication of the data.  For the evaluation of the system, we used LUBM and BSBM, two synthetic RDF/SPARQL benchmarks. Our experiments investigated the effect of workload-aware knowledge graph partitioning. The results showed a significant increase in the performance for the workload queries in comparison to a non-workload aware partitioning.

WawPart can easily be modified to function with other forms of knowledge graph representation such as graph databases, for example, Neo4j.  In the near future, we plan to test WawPart on a different knowledge graph system and investigate the impact of workload-aware partitioning on different types of knowledge graphs.

Furthermore, we are planning to extend WawPart to adaptive re-partitioning due the changes in the workload.  We plan to investigate the adaptability of partitioning due to

(1) the changes in the frequency of queries in the workload (currently, we assume that all queries are executed with the same frequency) and (2) the changes in the composition of the workload (queries may be eliminated and new queries may be added to the workload).